# Thermal hysteresis of spin reorientation at Morin transition in alkoxide derived hematite nanoparticles.

G. F. Goya [1]

*Instituto de Física, Universidade de São Paulo, CP 66318, 05315-970 São Paulo, Brazil*

M. Veith, R. Rapalavicuite, H. Shen and S. Mathur

*Institut für Neue Materialien, Saarland University Campus Im Stadtwald, D- 66041 Saarbruecken, Germany*

## ABSTRACT

We present results on structural and magnetic properties of highly crystalline $\alpha$-$Fe_2O_3$ nanoparticles of average size ~200 nm, synthesized from a novel sol-gel method using metal alkoxide precursor. These particles are multi-domain, showing the weak ferromagnetic-antiferromagnetic (WF-AF) transition (*i.e.*, the Morin transition) at $T_M$ = 256(2) K. Mössbauer measurements revealed a jump in hyperfine parameters (HP's) at $T \sim T_M$, which also displays thermal hysteresis upon cooling or heating the sample. The analysis of HP's as a function of temperature allowed us to discard temperature gradients as well as the coexistence of WF/AF phases as possible origins of this hysteretic behaviour. Instead, the hysteresis can be qualitatively explained by the small size and high-crystallinity of the particles, which hinder the nucleation of the WF or AF phases yielding metastable states beyond $T_M$.

PACS:   75.50.Tt (Fine-particle systems)

---

[1] *Corresponding Author: goya@unizar.es*

**INTRODUCTION**

The production of magnetic nanoparticles has been improved enormously in recent years, although the synthesis of high-purity nanoparticles of few nanometer size with controlled size and distribution is still a challenging problem [1-4]. In contrast to the solid-state processing, the chemical reactions in liquid phase allow a controlled interaction of atoms or molecules to form uniformly dispersed solid particles. The advantages of chemical processing can be augmented by assembling the phase-forming elements in molecular derivatives that transform into nanocrystalline ceramics under mild experimental condition ('*chimie douce*') [5,6]. This method allows a phase-selective synthesis, a key attribute for making oxides systems such as Fe-O, where several phases (hematite, maghemite, magnetite, etc) can be simultaneously formed in a narrow processing window.

Hematite, an antiferromagnet (AF) at low temperatures, displays a transition to a canted weak ferromagnet (WF) above $T_M \sim 260$ K. In the AF state, the magnetic moments are oriented along the [111] axis, and at $T_M$ there is a sudden flop to a new direction, parallel to the basal (111) plane, with a slight canting that yields a small net magnetization in this plane. This phenomenon was first observed experimentally 50 years ago [7], and explained by Dzyaloshinsky[8] and Moriya [9] as the consequence of an antisymmetric exchange interaction between atomic spins $\vec{S}$ of the form $\vec{D} \bullet (\vec{S}_i \times \vec{S}_j)$, where $\vec{D}$ is a constant vector.

In this letter, we describe a systematic investigation on the magnetic properties of highly crystalline $\alpha$-Fe$_2$O$_3$ particles of average size $\sim 200$ nm synthesized by the sol-gel processing of a single-source iron alkoxide. The size regime was chosen because it is just above range superparamagnetism of nanocrystalline materials (1-30 nm) and below the grain sizes of a microcrystalline material (several microns). This, in turn, allowed us to search for metastable WF/AF phases in highly crystalline small particles.

**EXPERIMENT**

The hematite particles analyzed here were produced by controlled hydrolysis of the alkoxide precursor [Fe(OBu$^t$)$_3$]$_2$, which is a single source for Fe and O. The xerogel obtained was calcined at 300 °C ( 6 h) to remove the organic residues and further heat-treated (1 h) at 400 °C to obtain pure hematite. Experimental details about the synthesis of precursor and the sol-gel processing can be found elsewhere.[10,11] The structure and morphology of the nanoparticles were examined by X-ray diffraction (XRD) and Transmission Electron Microscopy (TEM), respectively. Magnetization curves and hysteresis loops were recorded between 2 K ≤ T ≤ 350 K at different applied fields up to 30 kOe, using a commercial SQUID magnetometer. Mössbauer spectra were recorded in the transmission geometry between 78 K and 330 K using a 50 mCi $^{57}$Co/Rh-matrix source in constant acceleration mode. The estimated error in temperature (ΔT) was < 2 K. Two different cryostats (with He and N$_2$ as cryogenic liquids) were used to verify the absence of instrumental thermal hysteretic effects. The isomer shift are referred to α-Fe at room temperature.

**RESULTS AND DISCUSSION**

TEM images revealed a homogeneous distribution of highly facetted particles (fig. 1a) with an average particle size of ca. 200 nm. The high resolution image (inset fig. 1a) exhibits well defined lattice images of individual particles that confirmed their crystalline nature. This observation was also supported by the X-ray diffraction data, which revealed the nanocrystalline sample to be monophasic hematite (fig. 1b).

The magnetization as a function of temperature for different external fields is shown in fig. 2. We have found that, for applied fields up to 30 kOe, the Morin transition at $T_M$ = 256(2) K (defined as the inflection point of M(T) curves) does not change within the experimental accuracy. The width of the transition ($\Delta T_M$) extracted from the FWHM of the $d\chi_{dc}/dT$ curves was found to be < 12 K, and were also field-independent. Hysteresis curves at different temperatures (fig. 2, inset) show the appearance of a weak ferromagnetic moment and the concurrent rise of the coercive field $H_C$ above the transition to a WF state (T > $T_M$). The spontaneous magnetization value $\sigma_0$ = 0.19 emu/g, calculated from extrapolation of the high-field region of M(H) curve at T=350 K, is in agreement with the value expected from the small canted spin component in polycrystalline samples.[12]

Room temperature Mössbauer spectra showed a magnetically split sextet with hyperfine parameters close to the value found in bulk samples [13]. No evidence of superparamagnetic relaxation was observed up to 330 K, in agreement with the fact that the average particle size (~200 nm) observed from TEM measurements is above the critical size (~30 nm) for single-domain particles. The values of the line width $\Gamma$ = 0.28-0.29 mm/s (slightly larger than the calibration value of 0.28 mm/s) indicates high crystallinity of the investigated $Fe_2O_3$ particles.

As the temperature is decreased through $T_M$, the magnetic hyperfine field ($B_{hyp}$) and the electric field gradient (EFG) will sense the change in spin orientation from (nearly) perpendicular to parallel to the *c* axis. Fig. 3 shows the temperature dependence of hyperfine field ($B_{hyp}$), quadrupolar splitting (QS), isomer shift (IS), and linewidth ($\Gamma$) upon cooling and heating the sample through $T_M$. The most remarkable feature of these curves is the observation of a thermal hysteresis in hyperfine parameters when the sample crosses $T_M$ during heating or cooling. Although the width of this thermal hysteresis varies slightly for the four hyperfine parameters, it is well above the experimental error and can be estimated to be ~15 K. We have checked possible instrumental thermal hysteresis of the $N_2$-cryostat by doing independent measurements in a second, helium-cryostat at temperatures $T \sim T_M$ and the same results were obtained within experimental error. Some additional star-shaped points, displayed in fig. 3, correspond to this new heating process, which shows the accuracy of the data.

We have observed for the first time a small but measurable jump in IS values of ~ 0.03 mm/s, above the estimated experimental uncertainty (~ 0.01 mm/s). This jump is expected for the latent heat released at a first-order transition which modifies the temperature isomer shift. Fig. 3d also shows that the two maxima in the width of spectral lines ($\Gamma$) at $T_M$ match the hysteretic thermal response of the other hyperfine parameters. Possible explanations for the observed line broadening are: a) the coexistence of AF and WF phases across the transition and/or b) a softening of the spin structure that can cause a distribution in the relative orientation between the magnetic hyperfine field vector and the EFG tensor. If mechanism (a) is assumed, the expected broadening for a WF/AF two-phase spectra should be well above the maximum $\Gamma_{max} \sim$ 0.40 mm/s observed. Accordingly, attempts to fit the spectra at $T \sim T_M$ using two components were unsuccessful, indicating that the spectra are single-component along the whole temperature range. More important, since the broadening *also mirrors the thermal hysteresis* of QS and $B_{hyp}$ parameters (*i.e.*, the $\Gamma(T)$ curves also show hysteresis), it is clear that this increase of $\Gamma(T)$ at $T \sim T_M$ (see Fig. 3) cannot be originated from two phases (spectral components), since in that case no hysteresis effects in $\Gamma(T)$ should be observed. It is worth to note that Mössbauer spectra is the sum of individual atomic contributions, and thus the AF and WF phases will add independently, implying that the resulting spectra should display a maximum broadening at the same temperature independently of the heating/cooling direction. Thus, the observed hysteretic behaviour of $\Gamma(T)$ supports the idea that the more stable configuration extends beyond is equilibrium temperature, until some fast nucleation process is triggered. The above results are different from previous work on nanostructured hematite from Zysler *et al*. [14], where coexistence of AF and WF phases and absence of thermal hysteresis were reported.

Chow and Keffer [15] have demonstrated that the WF/AF transition can be triggered by a softening of surface magnons, assuming the existence of small perturbations to the molecular fields at the surface. The proposed mechanism involve the formation of surface areas of AF (or WF) 'nuclei' of soft magnons (where the spin rotation takes place), with a subsequent fast growth and propagation to the grain core when a certain critical temperature is reached. A consequence of this model is the appearance of hysteretic behaviour, by virtue of the different local surface/bulk anisotropies that allow soft magnons to operate at the transition in an opposed way depending on whether the system is cooled or heated.

The jump in QS values (fig. 3b) reflects the spin reorientation at $T_M$, through its dependence on the angle $\theta$ between the direction of hyperfine field $B_{hyp}$ and the principal axis of the EFG tensor. For $Fe_2O_3$, where the QS can be treated as a perturbation of the splitting from $B_{hyp}$, the shift $\varepsilon$ of excited state levels is given by [16]

$$\varepsilon = (-1)^{|m_I^*|+\frac{1}{2}} \left[ \frac{\Omega}{8} \left( 3\cos^2\theta - 1 \right) \right] \quad (1),$$

where $m_{I^*}$ are the magnetic quantum numbers of the excited state I* = 3/2, and $\Omega = e^2q_zQ$ reflects the interaction between the quadrupolar nuclear moment $Q$ of the probe atom and the EFG at the nucleus ($e^2q_z$). For the spin reorientation of $\Delta\theta$ = 90 degrees occurring at $T_M$, equation (1) leads to a ratio $\frac{QS(T<T_M)}{QS(T>T_M)} = -2$ which coincides, within experimental error with the observed jump from QS(T>$T_M$) = -0.21(1) mm/s to QS(T<$T_M$) = +0.39(1) mm/s. This experimental value further implies that no changes in the local symmetry are associated with the spin reorientation at $T_M$.

Previous work on hematite by van der Woude [17] using Mössbauer spectroscopy reported the observation of thermal hysteresis at T ~ $T_M$, although this work spanned a larger temperature range and consequently less attention was paid to this effect at T ~ $T_M$. Whereas the observed changes on hyperfine parameters can be explained on the basis of the Dzyaloshinsky-Moriya model, the thermal shift (hysteresis) observed when cooling/warming the samples could be connected with the metastable extension of one magnetic configuration beyond $T_M$, which seems to support the model of local field perturbations at the particle surface discussed above [15].

It has been proposed [18] that the H-T phase diagram of hematite should include a tricritical point ($H_{CR}$,$T_{CR}$), so the first- and second-order transition regimes are separated by critical lines. The abrupt change in $B_{hyp}$, QS and IS values observed experimentally in the present particles agrees with a first-order transition, and thus metastable states yielding hysteresis are likely to exist. [13]

In ref. [19], it has been proposed that crystal imperfections such as oxygen vacancies, dislocations, twinning, etc., can work as nucleation centers for the magnetic phases. In this line of reasoning, a metastable state of AF (WF) phase beyond $T_M$ on cooling (heating) could be observed only in nearly defect-free crystals (where nucleation is difficult), explaining why this effect has remained elusive.

Hysteretic behaviour of the Morin transition through M(T) and $B_{hyp}$(T) data has been also observed [20] in shock-modified $\alpha$-Fe$_2$O$_3$ particles with crystallite sizes < 200 nm. However, the samples were shock-treated and therefore resulted in particles with large accumulated crystal strain that could modify the magnetic properties (for example, the Morin transition was depressed by ~ 20-30 K). As the authors found clear evidence of coexistence of WF/AF phases at the same temperatures, it is likely that the origin of the hysteretic behaviour observed is the existence of two phases with different Morin temperatures. This could explain the observation of the effect in both magnetization and Mössbauer measurements.

For the present samples, the excellent crystallinity of the particles (as inferred from the small $\Gamma$ values) could prevent the nucleation process, yielding the observed hysteresis. The smooth, equilibrium heating/cooling conditions during Mössbauer measurements could also help to the stability. However, our magnetization data do not show any sign of hysteresis within experimental error ($\Delta T \sim 0.5$ K) even for the smallest applied field of 30 Oe. Since the AF/WF precursor 'nuclei' at the particle surface could be influenced by a moderate external magnetic field, we cannot discard the possibility that this hysteretic behaviour may be obliterated by even the smallest field (~ 30 Oe) used in our magnetization measurements. We note here that these fields are not the 'critical fields' required to 'spin flopping' the system, but to small energy differences between different surface regions, responsible of magnon softening at different temperatures. Further low-field magnetization and Mössbauer measurements should help to clarify this point.

In summary, we have shown that highly crystalline hematite samples with an average grain size of ca. 200 nm exhibit magnetic behaviour characteristic of bulk specimens. We have given clear experimental evidence of the existence of thermal hysteresis in both structural and magnetic hyperfine parameters at the Morin transition. The analysis of Mössbauer parameters showed that the spin reorientation is not accompanied by changes in crystal symmetry, and that there is no assessable coexistence of WF and AF phases at $T \sim T_M$. Instead, the broadening of the linewidth concurrently with the hysteretic effect suggests that magnon softening at the particle surface could be responsible for triggering the first-order phase transition.

SM and MV are thankful to the German Science Foundation for providing the financial support and a Visiting Fellowship to GFG in the framework of priority programme SFB-277 running at the Saarland University, Saarbruecken, Germany. GFG acknowledges partial support from Brazilian agencies FAPESP and CNPq under Grant Nos 01/02598-3 and 300569/00-9, respectively.

**REFERENCES**

1. S. Mathur, H. Shen, N. Lecerf, H. Fjellvåg and G. F. Goya, Adv. Mater. **14** (2002) 1405.

2. S. Mathur, in "Chemical Physics of Thin Film Deposition Processes for Micro- and Nano-Technologies", Y. Pauleau (ed.), NATO ASI – Kaunas 2001, Kluwer Academic Publ. 91, 2002.

3. J.S. Miller, Adv. Mater. **14** (2002) 1105.

4. V.S. Arunachalam and E.L. Fleischer, MRS Bulletin **26** (2001) 1020.

5. S. Mathur, M. Veith, H. Shen, S. Huefner and M. H. Jilavi *Chem. Mater.*14, 568, 2002.

6. S. Mathur, M. Veith, M. Haas, R. Haberkorn, H. P. Beck, H. Shen, S. Huefner *J. Am. Ceram. Soc*. 84(9), 1921, 2001.

7. F.J. Morin, Phys. Rev. **78**, 819 (1950).

8. I. Dzyaloshinsky, J. Phys. Chem. Solids **4**, 241 (1958).

9. T. Moriya, Phys. Rev. **120**, 91 (1960).

10. S. Mathur, M. Veith, V. Sivakov, H. Shen and H. B. Gao *J. Phys. IV* 11, 487, 2001.

11. S. Mathur, M. Veith, V. Sivakov, H. Shen, U. Hartmann and H. B. Gao, *Chem. Vapor. Depos.* **8** 277 (2002).

12. D. H. Anderson, Phys. Rev. **151**, 247 (1966).

13. A.H. Morrish, in "*Canted antiferromagnetism: Hematite*", World Scientific Publishing Co., Singapore, 1994.

14. R.D. Zysler, M. Vasquez-Mansilla, C. Arciprete, M. Dimitrijewits, D. Rodriguez-Sierra and C. Saragovi, Physica B **224** (2001) 39.

15. H. Chow and F. Keffer, Phys. Rev. B **10,** 243 (1974).

**Captions to the figures**

Figure 1. HR-TEM micrographs and XRD pattern of the $\alpha$-$Fe_2O_3$ particles.

Figure 2. Temperature-dependent susceptibility obtained at different external applied fields from 30 Oe to 30 kOe. Inset: M-H curves at T = 20, 200 and 350 K, showing the development of a WF magnetic moment above the Morin transition.

Figure 3. Mössbauer hyperfine parameters: (a) hyperfine field $B_{hyp}$, (b) Quadrupole splitting QS, (c) isomer shift IS and (d) linewidth $\Gamma$ measured in heating and cooling cycles. The solid star-shaped symbols are check points from the repeated measurements.

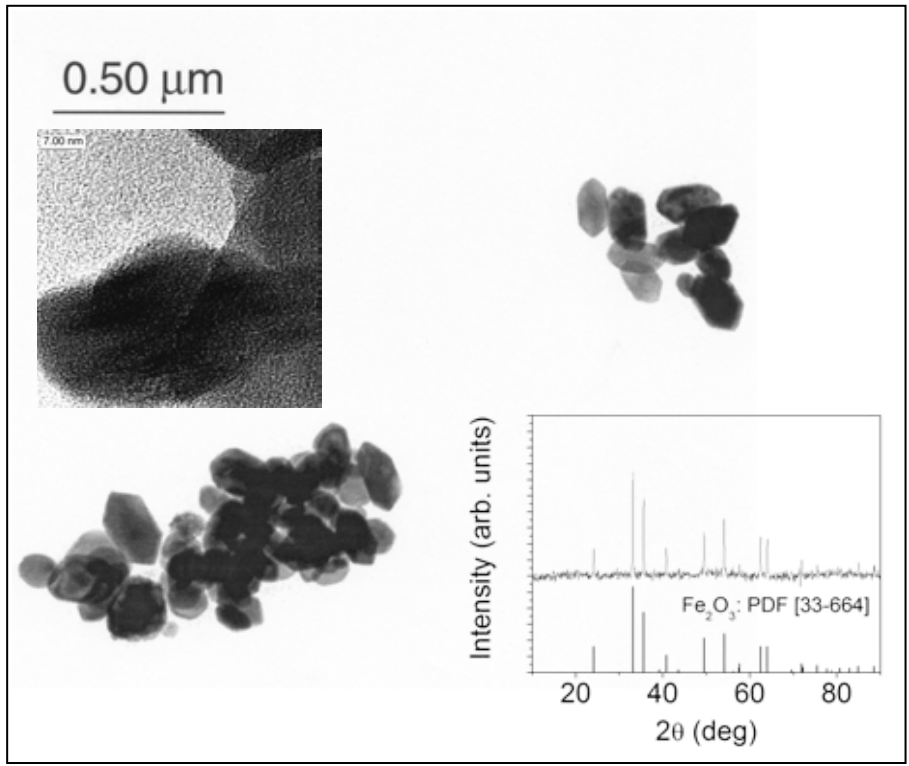

FIGURE 1

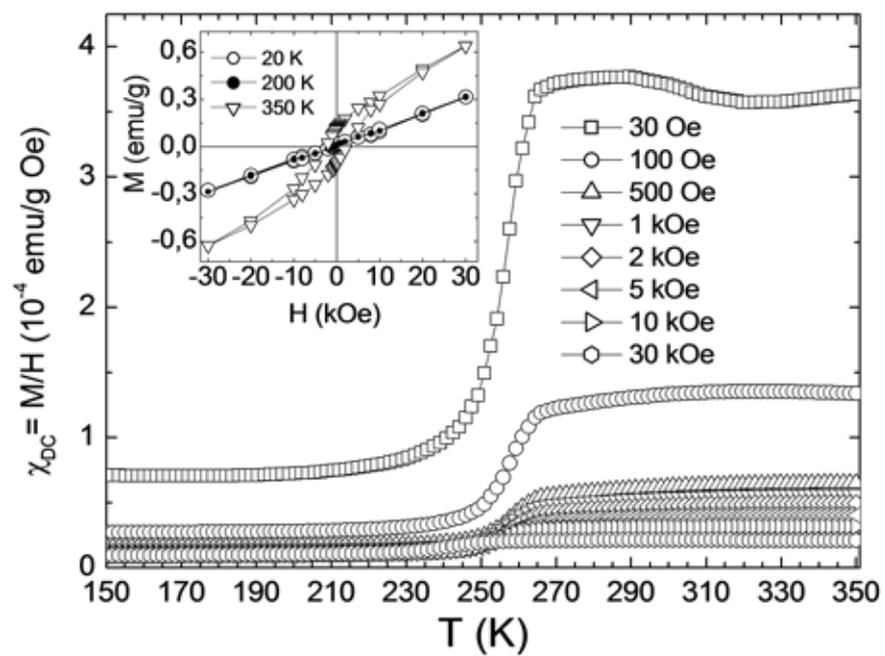

FIGURE 2

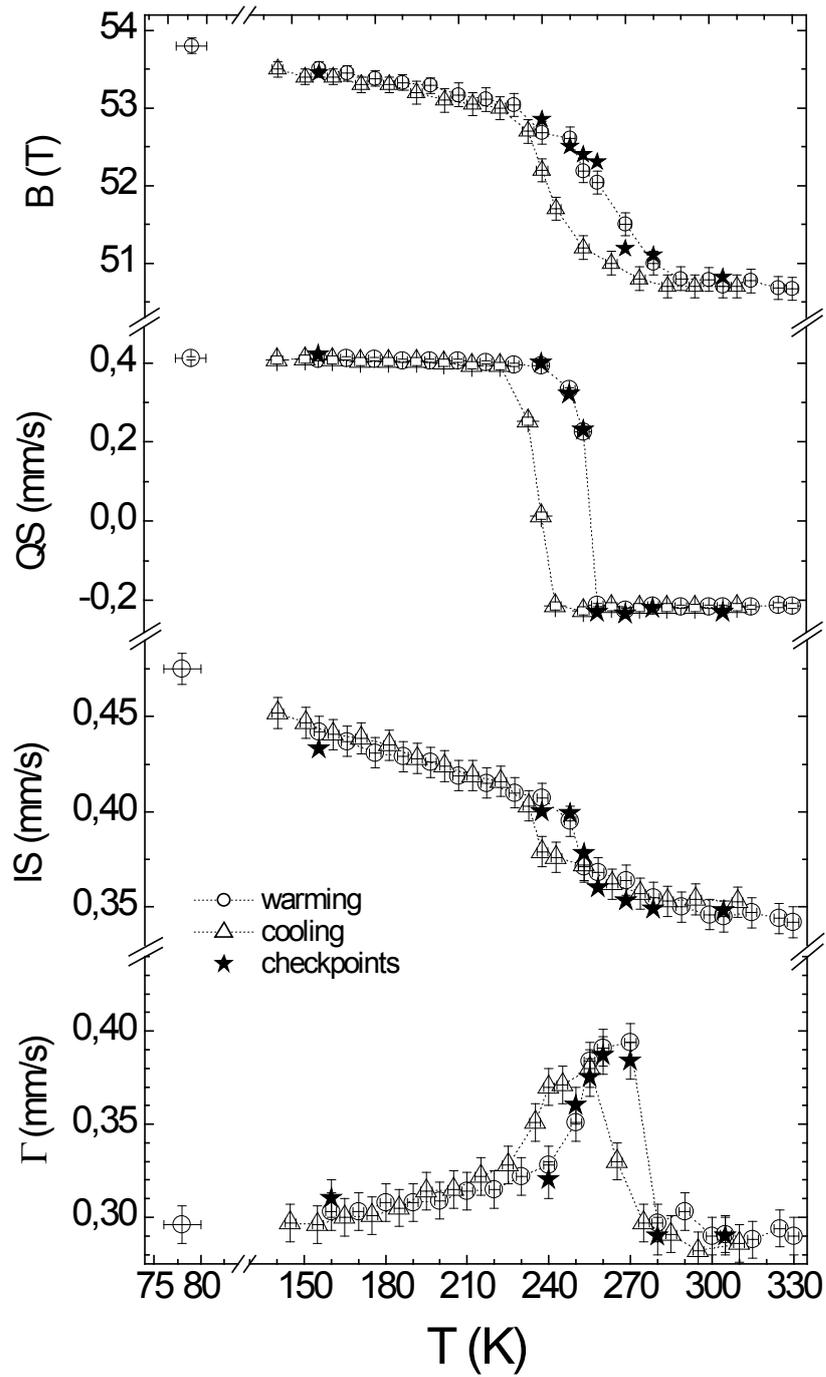